\documentstyle{article}
\newenvironment{namelist}[1]{%
\begin{list}{}
    {
     
     \settowidth{\labelwidth}{#1}
     \setlength{\leftmargin}{1.1\labelwidth}
    }
  }{%
\end{list}}
\begin{document}
\newcommand{\Ts}{T_{\rm sys}}
\newcommand{\Cel}{$^{\circ}$C}
\newcommand{\Fsp}{\vspace{0mm}}
% 
% --- Following are reference conventions used by Klwuer
%
\newcommand{\etal}{{\it et al.}}
\newcommand{\ApJ}{{\it Astrophys. J.}}
\newcommand{\MNRAS}{{\it Monthly Not. Roy. Astr. Soc.}}
\newcommand{\ASS}{{\it Astrophys. Space Sci.}}
\newcommand{\AaA}{{\it Astron. Astrophys.}}
\newcommand{\nat}{{\it Nature}}

%\begin{center}    % this forces two empty  spaces before the title
\begin{figure}[t]
\end{figure}

\begin{center}
{\bf THE GEM PROJECT: AN INTERNATIONAL COLLABORATION 
TO SURVEY
GALACTIC RADIATION EMISSION}\footnote{Presented by S. Torres
at the UN/ESA Workshop on Basic Space Sciences: 
From Small Telescopes to Space Missions, Colombo, 
Sri Lanka 11-13 
January 1996}  \\
\vspace{10mm}
S. TORRES\footnote{e-mail: 
storres@uniandes.edu.co},
V. CA\~{N}ON,
R. CASAS,
A. UMA\~{N}A \\
{\it Observatorio Astron\'{o}mico, Universidad Nacional de 
Colombia and Centro Internacional de F\'{\i}sica,
Bogot\'{a}, Colombia} \\ 
\vspace{4.0mm}    

C. TELLO, 
T. VILLELA \\
{\it INPE/CNPq, Brazil} \\
\vspace{4.0mm}    

M. BERSANELLI \\
{\it IFCTR-CNR, Milano, Italy} \\
\vspace{4.0mm}    

M. BENSADOUN, 
G. DE AMICI, 
M. LIMON,
G. SMOOT,
C. WITEBSKY \\
{\it Lawrence Berkeley National Laboratory, UC Berkeley, 
CA, USA} \\
\vspace{4.0mm}    
\end{center}

\vspace{20mm}

{\bf Abstract.} 
The GEM (Galactic Emission Mapping)
project is an international collaboration 
established
with the aim of surveying the full sky 
at long wavelengths with 
a multi-frequency radio telescope. 
A total of 745 hours of observation at 
408 MHz were completed from an Equatorial site 
in Colombia. 
The observations cover the celestial band
$0^h < \alpha < 24^h$, and 
$-24^{\circ} \ 22^{\prime} < \delta < 
+35^{\circ} \ 37^{\prime}$.
Preliminary results of this partial survey will be 
discussed. A review of the instrumental setup and a 
$\sim 10^{\circ}$ resolution 
sky map at 408 MHz is presented.

\section{The GEM Project}

Synchrotron radiation from relativistic electrons 
accelerated by the 
magnetic field of the Galaxy constitute
the main component of diffuse galactic 
emission at low frequencies (300 MHz to few GHz). 
At higher frequencies and high galactic latitude
free-free (bremsstrahlung) 
emission from ionized hydrogen starts becoming the dominant 
component (20 - 60 GHz) and beyond 60 GHz 
interstellar dust emission begins to dominate. 
A precise measurement and mapping of the diffuse galactic 
emission 
at low frequencies can tell us much about the cosmic ray 
electrons
and the dynamics of the Galaxy.

The discovery of cosmic microwave background
(CMB) anisotropies (Smoot \etal\ 1992) 
and subsequent measurements at large 
angle scales 
(Hancock \etal\ 1994; 
Ganga \etal\ 1993) have underlined the 
importance of an accurate determination 
of temperature spectral indices 
$(T_b \propto \nu^{-\beta})$
for the synchrotron and 
free-free  mechanisms in the Galaxy. 
The existing surveys
(Reich and Reich 1988;
see also  
Table 1 of Lawson \etal\ 1987 
for earlier work)
are insufficient to provide the 
required accuracy as pointed out by 
Davies \etal\ (1996). This 
fact is due to the large zero level 
and gain uncertainties
(i.e. $\pm 3$ K and 10\% for the 
408 MHz of Haslam \etal\ (1982)
and $\pm 0.5$ K and 0.5\% for the 
1420 MHz survey of Reich and Reich (1988)
respectively). 
Except for the Haslam \etal\ 408 MHz survey,
which is a composition 
of several patches made with four different 
telescopes, 
all others have partial sky coverage. 
The non-uniformity of these data sets and residual
striping effects constitute a serious limitations to
the quality of the maps. Accurate, multifrequency data
are needed in order to extrapolate the galactic emission
at higher frequencies 
(Brandt \etal\ 1994; Masi \etal\ 1991), 
especially due to 
the fact that the synchrotron
spectral index presents variations with 
galactic latitude (Lawson \etal\ 1987).
Future CMB experiments conducted 
from satellites such as 
COBRAS/SAMBA  and 
MAP  will carry
state of the art detectors but nevertheless
will be limited by the accuracy of
foreground emission removal.
There is a well justified need for a full sky,
homogeneous,
multi-frequency, 
and accurately calibrated 
survey of the galactic radio continuum.

Besides producing crucial information for CMB observations,
the GEM maps are scientifically important
in themselves.
The spectrum of
galactic radio emission at long wavelengths is 
dominated by synchrotron radiation emitted 
by relativistic cosmic ray electrons 
accelerated by the large scale galactic magnetic field. 
The synchrotron power emitted by electrons
depends on the electron energy density and 
the magnetic field intensity (Ginzburg \& Syrovatskii 1965). 
Thus, a survey 
of the radio emission provides useful information 
from which the galactic magnetic
field and the cosmic ray electron energy spectrum can 
be studied. Knowledge derived from these studies
have special relevance in models of cosmic ray 
acceleration.

Motivated by the above mentioned need to apply corrections 
to 
the galactic contamination present in cosmic microwave
background (CMB) maps, an international collaboration 
was established to measure the galactic
continuum emission in the range 408 - 5000 MHz
(De Amici \etal\ 1994).
The GEM 
collaboration was started by 
groups in 
Brazil (INPE/CNPq), 
Colombia (CIF and Observatorio Astron\'{o}mico), 
Italy (CNR), 
USA (Lawrence Berkeley National Lab/UCB),
and later joined by
Spain (IAC). 

\section{Experimental Setup}

One of the major difficulties faced by a full sky radio 
survey is the need to achieve accurate 
calibration. The design
strategy consists of a 
`portable' radio-telescope that can be moved to 
sites at different latitudes. Using the same calibrated
instrument allows for a consistent merging of patches 
of the sky taken at different sites. 
An additional advantage is that 
moving the instrument at different latitudes allows
pointing the main beam at a small angle 
from the local zenith, thus
minimizing the atmospheric absorption seen at large
zenith angles (important at the higher frequencies). 
The GEM telescope is mounted 
on a rotating base and the pointing of the main beam 
is kept at a fixed angle from the zenith
to keep the atmospheric contribution constant.
The mechanical design of the mount system 
allows for changes in the zenith angle of the antenna. 
The antenna sidelobes can be checked by
observing the same part of the sky at different zenith 
angles,  therefore 
obtaining a direct measurement of ground contribution.
The combined motion of the rotating base and the Earth's 
rotation results in a swath of the sky seen at each latitude. 
In principle one could cover the whole sky by collecting 
data in this mode from three equidistant latitudes 
(i.e. $60^{\circ}$S, $0^{\circ}$ and $60^{\circ}$N).
In practice it is desirable to allow for overlaps 
of the covered regions so as 
to ensure a self-consistent
data set. We have acquired several hundred 
hours of observation at 408, 1465, 2300 and 5000 MHz
from an Equatorial site in Colombia and 
a Northern site in Bishop, California. 
The telescope is currently at the IAC-Tenerife Observatory
in Spain. An overview of the experimental setup and 
preliminary results of the 408 MHz survey conducted from 
the Equatorial site will be presented.

A block diagram (Fig.~1) shows the main 
parts that form the GEM system.  
The main section is made of the parabolic 
reflector, the feed antenna, the rotating 
base and 
the radiometer. Table~\ref{tble:param} summarizes 
the instrumental parameters of GEM some of which 
have been computed as described below. 

\begin{table} [htb]
\caption{GEM parameters} \label{tble:param}
\begin{center}
\begin{tabular}{ll} \hline
Parameter                &   value \\ \hline \hline
System \\
Receiver frequencies    & 408, 1465, 2300, 5000 MHz \\
Reflector diameter              &  5.5 m  \\ 
Reflector diameter with extension & 9.5 m \\
Postdetection integration time    & 0.56 s  \\
Base rotation speed (nominal)     & 1 rpm   \\
\hline
408 MHz Receiver         &   \\
System temperature       & $104 \pm 6$ K \\
Gain                     & $58 \pm 1$ K V$^{-1}$ \\
Band width               & 28 MHz \\
Sensitivity              & 26 mK/integration time \\
Beam width (FWHM)        & $11.3^{\circ}$ \\
Baseline susceptibility  & $-3$ K \Cel $^{-1}$ \\
Gain susceptibility  & $-2.7 \times 10^{-4}$ \% \Cel $^{-1}$ \\
\hline
\end{tabular} 
\end{center}
\end{table}

\subsection{Parabolic reflector, 
base and feed antenna}

GEM uses a Scientific Atlanta  5.5-m parabolic 
reflector mounted on an alt-azimuth rotating 
base. 
The 408, 1465 and 2300 MHz receivers use a prime focus feed: 
a backfire helix 
at the low frequencies, and a conical antenna at 2300 MHz.
The 5000 MHz receiver unit and conical antennae
are mounted at the Gregorian focus. 
Aluminum panels extend the parabolic reflector 
surface to a total diameter of 9.5-m. The purpose 
of this aluminum shield is to minimize diffracted
ground emission
and to
allow us to determine the beam efficiency and loss by
covering up with highly reflective opaque 
material half  the $4\pi$ solid angle.

The back-fire helix feed antenna (408 MHz) 
consists of a 9.5 turns 
made of copper piping of 9.6 mm
diameter. The turn length is $0.92 \lambda$,
the spacing between turns is 15.4 cm and the axial length  
is 148 cm.  
The feed antenna is sensitive to circularly polarized 
radiation.
The main lobe width of the combined 
antenna/reflector  assembly
is obtained using the transit of the sun
in front of GEM. 
A plot of the 408 MHz signal voltage versus the separation
angle between GEM's pointing and the sun shows 
the Gaussian-like shape of the main lobe with 
FWHM of $11.3^{\circ}$.

The mount assembly rests on a rotating 
base with a  velocity of 1 rpm.
An azimuth angle encoder mounted on the 
rotating axis of the main
GEM assembly and
a similar encoder on the horizontal axis 
provide
a 0-10 V analog voltage proportional to the angle.
The zero angle 
resulting from this reading is calibrated 
using the sun signal in the data in combination 
with the sun ephemerides.  These calibration parameters
were verified by checking GEM's mechanical orientation 
with respect to the geographic North employing a 
theodolite.

\subsection{The 408 MHz receiver}

The 408 MHz radiometer uses a total power receiver 
with two RF amplification stages and one DC amplification 
($\times 1000$) after detection (Fig.~2). 
A cavity filter 
($\Delta \nu = 28$ MHz) at the front end of the receiver and 
a tubular filter after RF amplification  
are used. 

The ambient temperature of the receiver 
is controlled and isolated 
from the outside temperature by 
warming the inside of the receiver box 
with resistance heaters while cooling 
the inside of a bigger box that
encloses the receiver box. 
Temperature sensors and a regulating circuit
keep the  operating temperature 
of the receiver
to within $\pm 0.2$ \Cel.

The susceptibility of the radiometer output voltage 
baseline to changes on operation temperature 
is measured as
$-3$ ${\rm K}/^{\circ}C$.
Thus, within the temperature stability 
achieved one expects baseline drifts up to 
$0.6$ K which typically occur in time scales
of 8 to 9 hours. 
We correct for temperature 
induced baseline drifts by using the 
information from the various temperature
sensors and the ${\rm K}/^{\circ}C$ slope
quoted above.
Gain changes are monitored 
by injecting a fixed amplitude reference pulse 
every 45 seconds. The reference noise pulse is 
generated by a diode and is connected to the
input of the first
amplifier stage via a directional coupler.

Radio frequency interference (RFI), due to 
local radio communications or electric 
discharges in the atmosphere can cause 
excessive dispersion of the observed signal.
An RFI detection circuit signals the presence 
of such anomalous data.
This circuit 
produces an output voltage $V_{\rm sat}$ proportional 
to the number of times that the signal average 
crosses some preset threshold during the integration time. 
Cutting data above some $V_{\rm sat}$ threshold effectively
acts as a low-pass filter on the signal.

\subsection{Gain and system temperature calibration}

Gain, $G$, and system temperature, $\Ts$, are 
obtained by recording the radiometer signal voltage
as the input is switched from a cold to a hot
target. 
A $50 \Omega$ termination submerged in a 
bath of liquid Nitrogen (LN) at ambient pressure
is used as a reference cold target. 
A similar termination at room temperature
provides the warm reference target. 
The temperature of the warm target is monitored by 
a thermocouple while that of the cold target, 
75 K,
is measured using a reference
platinum resistance immersed in the LN.
Including the contributions of the connectors and 
the attenuation and reflection properties
of the coaxial
cable (at 408 MHz) connecting the termination 
to the radiometer brings the equivalent
LN temperature to $92 \pm 2$ K.  
The measured parameters are:
$G = 58 \pm 0.9 \pm 0.2$ K V$^{-1}$, 
$\Ts = 104 \pm 6 \pm 1$ K. 
The first error is systematic, 
the second statistical.  
Systematic 
errors are obtained with the error propagation 
formulas  taking into account 
the errors in the measurements of 
target temperatures and signal 
voltages. 
The statistical errors quoted come 
from the standard deviation of the 
measurements done over a period of 1 hour.
The sensitivity of the receiver implied
by the measurement of $\Ts$ is 
26 mK/integration time
and is consistent with the {\it rms}
spread of the antenna temperature when looking 
at a fixed and known temperature target. 

\subsection{Data acquisition system}

A 16 bit ADC 
samples 14 multiplexed analog channels
(radiometer signal, 
angle encoders,
temperature sensors, 
$V_{\rm sat}$ and
noise source voltage)
and places the digitized information 
on a serial output. 
A time stamp and a sequential identifier number
(`frame number') is added to the data stream.
The time tag is the UT time provided by an 
external digital clock/receiver synchronized with 
the WWV station in Denver. 
Pulses from an internal clock (100 Hz) are 
divided down to provide the integration time
of 0.56 seconds, which effectively defines the 
basic data rate of 50 bytes/s.

The data acquisition
(DAQ) unit is mounted in a separate NIM crate box 
near the receiver whose DC output signal
is connected 
via coaxial cable. 
The output of the DAQ unit is 
serial digital data sent to the 
DAQ computer in the control room 
through slip rings.
The DAQ computer (a Macintosh Performa) 
performs a routine data quality check 
and stores the data on disk for future 
analysis.

\section{The Equatorial Site}

To complement the observations made or planned 
from California,  Brazil and Spain 
an Equatorial site seemed to be the natural
place to look. Colombia, in particular, is located 
on the Equator and has high peaks. 
However, the tropical 
weather conditions characteristic of equatorial
latitudes can place severe restrictions.
Site studies (Torres \etal\ 1992; Hoeneisen \etal\ 1992)
in Colombia showed that there exist
at least two dry seasons suitable for 
radio-astronomy  and that there is a large
variation of weather conditions at the different
regions of the country. Due to the complex terrain 
there are micro-climates with very dry regions. 
The selected site in
Villa de Leyva  (Boyac\'{a}, Colombia) 
is a region 
characteristic for its dry atmosphere, specially
during the dry season (December - April). 

The GPS coordinates and altitude of the site are: 
LAT $=  5^{\circ} \, 37^{\prime} \,
7.84^{\prime\prime} \pm 0.59^{\prime\prime}$ N,
LONG $=   
73^{\circ} \, 35^{\prime} \,
0.53^{\prime\prime} \pm 0.72^{\prime\prime}$ W,
Altitude $=  2,173 \pm 28$  m.a.s.l.
The errors quoted are statistical while the
systematic error is the $3.2^{\prime\prime}$ allowed by DOD.

\section{Data Reduction}

Every $0.56$ seconds the DAQ module sends via 
serial RS-232 a 14 words frame of data to the on-line
computer.
The off-line analysis software prepares the data
for pixelization and
applies cuts as follows:
azimuth and elevation angles are calibrated;
Julian and Sideral time of each measurement is computed; 
data points with the sun  within $30^{\circ}$ of the 
beam are rejected;
noise source pulses are subtracted from the data  
and used for gain corrections;
strong RFI signals are extracted by binning short segments 
of data 
in azimuth and rejecting data that shows large deviations 
from  the mean of its corresponding bin;
a polynomial fit baseline is found and corrected for 
temperature induced drifts; finally,
the radiometer signal is  calibrated and pixelized.
Further cleaning of the data is done on the pixelized 
set by rejecting points that show large deviations
with respect to the mean temperature of their corresponding
pixel.

A total of 1,116 hours of observation at 408 MHz were 
completed 
from the Equatorial site from March 13 to May 13 (1995). 
Data runs with strong presence of RFI 
were rejected. Data
with elevation angles different than 
$60^{\circ}$ were used for ground contribution studies.
The remaining
745 hours of data were analyzed.
From the selected data-set 
40\% is rejected as follows:
10.6\% is sun contaminated,
6.3\% is RFI contaminated,
3.8\% is labeled `anomalous' and is cut at the time of the 
baseline fit,
6.4\% due to temperature sensors out of bounds,
3.5\% RFI contaminated as signaled by $V_{\rm sat}$,
and the remaining goes into the other cuts.

\section{The GEM 408 MHz sky map}

As data-base structure to store pixelized data
we used the {\it skycube} pixelization scheme used 
by COBE
(Torres \etal\ 1989, Torres 1995). 
This choice is dictated by 
the availability of
several libraries and procedures
to handle sky maps in this format
and because it makes it easier to compare our 
maps with others. 
With the 408 MHz beam size  of 
$\sigma_{\rm beam} \sim 8.38 \times 10^{-2}$ radians
GEM can distinguish 
$\sim (2/\sigma_{\rm beam})^2 \approx 600$ independent 
pixels in the sky.  
Thus, {\it skycube}
pixelization level 5 with 1536 pixels seems appropriate.

The primary output of the analysis software is 
a sky map with
a 54\% sky coverage corresponding to the 
$60^{\circ}$ wide celestial band to which we have 
access from the Equatorial site. This band 
covers the sky region 
$0^h < \alpha < 24^h$, 
$-24^{\circ} \ 22^{\prime} < \delta < 
+35^{\circ} \ 37^{\prime}$
in right ascension and declination respectively.  
Fig.~3 is a rendering of this map in
an equatorial celestial projection. 
The number of observations per pixel goes from 18 to 
13,266 with a mean of 3,576 and a sigma of 2,188.

The standard deviation of the temperature
for each pixel is $\sim 2$ K, which is 
somewhat higher than expected from 
the measured sensitivity, indicating that 
there remains spurious effects not completely 
removed from the signal. 

The zero level of the calibrated map needs 
corrections due to all possible known contributions
not originating in the Galaxy.
The integrated background of faint extragalactic
background radio sources at 408 MHz is 3.19 K
(Lawson \etal\ 1987).
The CMB temperature is $2.726 \pm 0.010$ K 
(Mather \etal\ 1994), 
atmospheric  emission at 408 MHz is $\leq 1$ K.
The contribution due to antenna and transmission line
insertion loss and sidelobes pickup
is estimated using data at several elevations and 
comparing data taken during the day with data 
taken at night
(difference in ambient temperature $\sim 20$ $^{\circ}C$).
The combined contribution of all these factors 
amounts to $17 \pm 8$ K and has been taken into account 
in the map in Fig.~3. 
The preliminary data  presented here 
still has larger uncertainties than 
what is required for CMB foreground removal.
Work towards a more 
accurate determination of the beam pattern, 
antenna insertion loss and calibration 
parameters is in progress. 
A preliminary comparison with the map 
of Haslam \etal\ was done by means of
a pixel-to-pixel correlation between the GEM 
map and the same region of the sky in the Haslam 
\etal\ map degraded to the same 
angular resolution as GEM.
The correlation 
coefficient is 98.3\% 
which is highly significant. 
A more detailed comparison will be 
treated in a forthcoming paper.

The tolerable errors in
long-wavelength surveys used in CMB work
are quite stringent (Brandt \etal\ 1994).
This fact is specially relevant when the
`subtraction technique' is used (i.e. 
measure the foregrounds in the 
region of the spectrum where they are strong, 
and extrapolate to the millimeter and 
sub-millimeter region where the CMB is measured). 
The rapid propagation of errors in the interpolation
does not favor this method. However, a multi-frequency 
analysis (Masi \etal\ 1991) or a multi-frequency 
Wiener filtering analysis (Tegmark \& Efstathiou 1995)
seem to be a more promising choice to remove foregrounds
from CMB maps. Thus GEM multi-frequency data in the 
range 408 - 5000 MHz combined with 
IRAS 100 $\mu$m,
COBE-DMR (31.5, 53 and 90 GHz) and 
COBE-DIRBE 140 $\mu$m data  
should be sufficient to separate 
the galactic foreground from  CMB maps.

{\bf Acknowledgments.} --- 
This work was supported in part
by the Director, Office of Energy Research, 
Office of High Energy and 
Nuclear Physics, 
Division of High Energy Physics of he U.S. Department
of Energy under contract No. DE-AC03-76SF00098,
the National Science Foundation 
under the
International Joint Research NSF 9102295,
and Calspace grant No. CS-63-94.
Colciencias (projects 2228-05-007-92,
and 2228-05-103-96) 
provided funds
to support GEM operations in Colombia.
We owe thanks to many persons who helped 
make this a succesful project. In particular
we would like to thank 
H. Arias,
L. M. Cruz, 
M. Fonsesca, 
J. Gibson, 
D. Heine, 
G. Holgu\'{\i}n
J. Larsen,
D. Maino, 
E. Mattaini,
A. Pantoja, 
L. Pe\~{n}a,
E. Santambrogio,
J. Yamada.

%\newpage

\begin{flushleft}
{\bf REFERENCES}
\end{flushleft}

\begin{namelist}{xxx}
\setlength{\parskip}{0.0mm}
\setlength{\parsep}{0.0mm}
\setlength{\itemsep}{0.0mm}
\setlength{\labelsep}{0.0mm}
\raggedbottom
\item[Brandt], W. N. \etal : 1994, \ApJ\ {\bf 424}, 1.
\item[Davies], R. D., \etal : 1996, \MNRAS\ {\bf 278}, 925.
\item[De Amici], G., \etal : 1994, \ASS\ {\bf 214}, 151.
\item[Ganga], K., \etal : 1993, \ApJ\ {\bf 410}, L57.
\item[Ginzburg], V. L. \&  Syrovatskii, S. I.: 1965, 
  {\it Annu. Rev. Astron.  Astrophys.} {\bf 3}, 297.
\item[Hancock], S. \etal : 1994, \nat\ {\bf 367}, 333.
\item[Haslam C.G.T], \etal : 1982, 
 {\it Astr. Astroph. Suppl. Ser.} {\bf 47}, 1.
\item[Hoeneisen], B. \etal : 1992, {\it Nuclear Phys. B 
    (Proc. Suppl.)}, {\bf 28B}, 191.
\item[Lawson], K. D., \etal : 1987, \MNRAS\ {\bf 225}, 307.
\item[Masi], S., \etal : 1991, \ApJ\ {\bf 366}, L51.
\item[Mather], J. C., \etal : 1994, \ApJ\ {\bf 420}, 439.
\item[Reich], P. and Reich W.: 1988, \AaA\ {\bf 74}, 7.
\item[Smoot], G. F., \etal : 1992, \ApJ\ {\bf 396}, L1.
\item[Tegmark], M. \& Efstathiou, G.: 1995, MPI-PhT/95-62 
  (preprint).
\item[Torres], S., \etal : 1989, in 
  {\it Data Analysis in Astronomy III}
  (eds V. di Gesu, L. Scarsi, and  M.C. Maccarone) 
  {\bf 40}, 319  (New York: Plenum)
\item[Torres], S.: 1995, \ASS, {\bf 228}, 309.
\item[Torres], S., \etal : 1993, 
  {\it Rev. Col. de F\'{\i}sica}, {\bf 25}, 23.
\end{namelist}
%
% - - - -
%
%
%\newpage
\vspace{20mm} 
\begin{flushleft}
{\bf FIGURE CAPTIONS}
\end{flushleft}

\begin{namelist}{xxxxxxxxxx}
\setlength{\parskip}{0.0mm}
\setlength{\parsep}{0.0mm}
\setlength{\itemsep}{0.0mm}
\setlength{\labelsep}{0.0mm}
\raggedbottom

\item[Figure 1.]~Block diagram of the GEM system.

\item[Figure 2.]~Diagram of the 408 MHz receiver.

\item[Figure 3.]~GEM 408 MHz sky map contour levels
in equatorial celestial coordinates. \\
The temperature levels go from 22 to 160 K in 
constant increments of 4 K up to 100 K where the 
increment changes to 20 K.
\end{namelist}
\end{document}